\documentclass{article}
\usepackage{spconf,amsmath,graphicx}
\usepackage{fancyhdr}
 \usepackage{epsfig,amssymb,amsmath,multirow,boldline,graphicx,makecell,hyperref,subfig}
\usepackage{kotex}
\usepackage{float}

\usepackage{xcolor}

\title{Domain attentive fusion for End-to-End Dialect Identification\\ with unknown target domain}
\name{ Suwon Shon$^1$, Ahmed Ali$^2$, James Glass$^1$}

\address{MIT Computer Science and Artificial Intelligence Laboratory, Cambridge, MA, USA$^1$  \\
Qatar Computing Research Institute, HBKU, Doha, Qatar$^2$\\
{\small \tt \{swshon,glass\}@mit.edu \qquad amali@qf.org.qa} }
\begin{document}
\ninept
\maketitle
\begin{abstract}
End-to-end deep learning language or dialect identification systems operate on the spectrogram or other acoustic feature and directly generate identification scores for each class. 
An important issue for end-to-end systems is to have some knowledge of the application domain, because the system can be vulnerable to use cases that were not seen in the training phase; such a scenario is often referred to as a domain mismatched condition.   In general, we assume that there is enough variation in the training dataset to expose the system to multiple domains.   In this work, we study how to best make use a training dataset in order to have maximum effectiveness on unknown target domains.  Our goal is to process the input without any knowledge of the target domain while preserving robust performance on other domains as well. 
To accomplish this objective, we propose a domain attentive fusion approach for end-to-end dialect/language identification systems.  To help with experimentation, we collect a dataset from three different domains, and create experimental protocols for a domain mismatched condition.  The results of our proposed approach, which were tested on a variety of broadcast and YouTube data, shows significant performance gain compared to traditional approaches, even without any prior target domain information.

\end{abstract}
\begin{keywords}
Dialect identification, language identification, self-attention, fusion
\end{keywords}
\section{Introduction}
\label{sec:intro}
Channel or domain mismatch between training and test data can be a significant factor affecting performance for language and dialect identification (DID) systems, but mismatch has not been addressed as seriously for these tasks as it has been in the speaker recognition arena. In 2013, a domain adaptation challenge (DAC13) was held on domain mismatch for speaker recognition~\cite{dac13}. From the success of DAC13, many researchers explored the domain mismatch problem on the speaker recognition task~\cite{Aronowitz2014a,Garcia-Romero2014b,Villalba2014,Shon2017}.  However, the same mismatch issue for language/dialect recognition was not actively studied until the NIST 2017 Language Recognition Evaluation (LRE)~\cite{lre17} provided speech datasets from multiple domains.  At both challenges, many studies tried to adapt the Gaussian Back-end or PLDA back-end on top of the i-vector or x-vector speaker embeddings~\cite{Mclaren2018domain, Lopez2018domain, Aronowitz2014a, Shum2014, Garcia-Romero2014b, Villalba2014}.  Although these approaches cannot be directly applied to end-to-end deep learning systems for these same tasks, they achieved reasonable performance when the target speech domain was known {\it a priori}. 

For dialect identification task, the Multi-Genre Broadcast 3 (MGB-3) challenge also provided domain mismatched data. Unsupervised learning of dialectal speech was investigated by Zhang~\cite{zhang2018language} and Shon~\cite{Shon2018unsuper,Shon2017c} to extract domain invariant features from MGB-3 dataset. By exploiting speech data from several domains without explicit language and domain labels, the networks could extract domain invariant representations from input speech. The approaches still needed some amount of labeled data to train subsequent identification systems.  They achieved large performance gains when there were no language labels on the target domain training dataset compared to traditional acoustic features like MFCCs. Although the performance gap closed when enough labeled target domain data were available, they have an advantage for scenarios where large amounts of unannotated speech is available~\cite{Shon2018unsuper}.

In this research, we do not assume any resource limitation or challenging situations like unlabeled target domain data.  Instead we assume that we have enough data from multiple domains with labels for dialect identification.  However, we also assume that we don't have any domain information about the target speech.  In this case, a training model with labeled multiple domain data would easily provide superior performance over the previous efforts which adapt the back-end scoring to a target domain.  Another possible approach is that score-level fusion of subsystems which are trained on single domain data. In the periodic series of NIST evaluations, it was observed that linear fusion of multiple subsystems consistently outperforms the single best system~\cite{Lee2017}.  However, the performance of the fusion system depends strongly on the logistic regression fusion, whose parameters need to be calibrated to specific trials which reflect the test conditions. Thus, the system fusion was optimized to the specific domain of the test trials, so that if the test speech came from a random domain, the fusion system cannot guarantee the best performance.

To address the unknown domain speech input, we propose to use a self-attention layer in our end-to-end model and have fusion parameters which are calculated from the input speech. Once the domain attentive layer is trained using the training data, it automatically generates the best fusion weight of domain-specific systems by taking the output of each subsystem. Thus, ideally, the optimal fusion weight would be generated for every single input. 

In the following sections, we examine baseline systems for unknown domain inputs and propose domain attentive layers.  We also describe our data collection from YouTube, called Varieties and Dialects (VarDial) 2018, to provide a dataset for our experiments.

\begin{table*}[t]
\centering
\resizebox{0.7\textwidth}{!}{%
\begin{tabular}{r|r|r|r|r|r|r|rrrr}
\hlineB{2}
\multicolumn{1}{l|}{Data name} & \multicolumn{6}{c|}{MGB-3} & \multicolumn{4}{c}{VarDial 2018} \\ \hline
\multicolumn{1}{l|}{Type} & \multicolumn{2}{c|}{Training} & \multicolumn{2}{c|}{Development} & \multicolumn{2}{c|}{Testing} & \multicolumn{2}{c|}{Training} & \multicolumn{2}{c}{Testing} \\ \hline
\multicolumn{1}{l|}{Domain} & \multicolumn{2}{c|}{Recorded Broadcast} & \multicolumn{4}{c|}{High-quality Broadcast} & \multicolumn{4}{c}{YouTube} \\ \hline\hline
\multicolumn{1}{c|}{Dialect} & \multicolumn{1}{c|}{Ex.} & \multicolumn{1}{c|}{Dur.} & \multicolumn{1}{c|}{Ex.} & \multicolumn{1}{c|}{Dur.} & \multicolumn{1}{c|}{Ex.} & \multicolumn{1}{c|}{Dur.} & \multicolumn{1}{c|}{Ex.} & \multicolumn{1}{c|}{Dur.} & \multicolumn{1}{c|}{Ex.} & \multicolumn{1}{c}{Dur.} \\ \hlineB{2}
EGY & 3,093 & 12.4 & 298 & 2.0 & 302 & 2.0 & \multicolumn{1}{r|}{93,408} & \multicolumn{1}{r|}{206.3} & \multicolumn{1}{r|}{1,143} & 5.5 \\
GLF & 2,744 & 10.0 & 264 & 2.0 & 250 & 2.1 & \multicolumn{1}{r|}{92,603} & \multicolumn{1}{r|}{204.5} & \multicolumn{1}{r|}{1,147} & 5.6 \\
LEV & 2,851 & 10.3 & 330 & 2.0 & 334 & 2.0 & \multicolumn{1}{r|}{232,585} & \multicolumn{1}{r|}{513.6} & \multicolumn{1}{r|}{1,131} & 5.5 \\
MSA & 2,183 & 10.4 & 281 & 2.0 & 262 & 1.9 & \multicolumn{1}{r|}{9,518} & \multicolumn{1}{r|}{21.0} & \multicolumn{1}{r|}{944} & 4.6 \\
NOR & 2,954 & 10.5 & 351 & 2.0 & 344 & 2.1 & \multicolumn{1}{r|}{24,841} & \multicolumn{1}{r|}{54.9} & \multicolumn{1}{r|}{980} & 4.8 \\ \hline
Total & 13,825 & 53.6 & 1,524 & 10.0 & 1,492 & 10.1 & \multicolumn{1}{r|}{452,955} & \multicolumn{1}{r|}{1000.3} & \multicolumn{1}{r|}{5,345} & 26.0 \\ \hlineB{2}
\end{tabular}%
}
\caption{Arabic dialect data breakdown for the MGB-3 and VarDial 2018 datasets.}
\label{tab:data}
\end{table*}

\section{Dialectal language dataset}
For this work, we used the two dialect datasets called MGB-3 and VarDial 2018, to generate domain mismatched conditions for our experiments.   As shown in Table~\ref{tab:data}, the MGB-3 data consists of recorded and high-quality broadcasts, while the VarDial data consists of YouTube videos.  Each dataset contains data that has been labeled from five Arabic dialects: Egyptian (EGY), Levantine (LEV), Gulf (GLF), North African (NOR), and Modern Standard Arabic (MSA).  The MGB-3 data collection was distributed equally across all five dialects for both training/dev/test partitions, while the VarDial data has significantly more data, though more unevenly spread across dialects, due to the manner in which it was collected.  Although the MGB-3 development set is relatively small compared to the training set, it matches the test set channel conditions, and thus provides valuable information about the test domain.  More details about MGB-3 are available in ~\cite{Ali2017}.

The VarDial 2018 dataset was collected from YouTube in a semi-supervised technique. Initially, we identified more than 30 YouTube channels. The dialect for each channel is known. However, we are unable to guarantee that there is no cross-dialectal speech in the channels. For every channel, we crawled more than 100 video clips. For each video, we ran voice activation detection~\cite{meignier2010lium} and the data was sliced into small audio clips between 5 and 30 seconds. Furthermore, the test set was manually verified and was uniformly distributed. The accuracy of verifying the test set and the non-speech clips were about 92\%. More details about VarDial 2018 are available in~\cite{Zampieri2018}.

\section{Baseline System DID Experiments}

\subsection{End-to-end dialect identification system}
In this work, we adopt the end-to-end dialect identification system proposed in~\cite{Shon2018odyssey}.
This system has a stack of convolutional neural network (CNN) layers, followed by a global pooling layer that aggregates frame level representations to produce utterance level representations. The output of the global pooling layer is followed by two feed forward (FF) layers. Specifically, the network consists of four one-dimensional CNN layers (40$\times$5 - 500$\times$7 - 500$\times$1 - 500$\times$1 filter sizes; with 1-2-1-1 strides; the number of filters is 500-500-500-3000) and two FF layers (1500-600). The size of the final softmax layer is determined by the task-specific language or dialect labels and the softmax output can be used directly as a score for each dialect class for the DID task. We used MFCCs as inputs to the end-to-end system since they obtained the best performance without any dataset augmentation.  Note that no dataset augmentation was performed for these experiments.

\subsection{Training on multiple domains}

Consider datasets $S_1$ and $S_2$ with two unknown data distributions $\mathcal{D}_1$ and $\mathcal{D}_2$. If the target domain is the same as $S_1$, we can discard $S_2$ and use only $S_1$ to train a network. To cope with multiple domains, multiple networks could be learned using each domain dataset. In this case, we need $N$ networks for $N$ domain target domains.

Score-level fusion of $N$ single-domain networks can boost performance. Logistic regression based fusion is a common method for learning an optimal linear combination of the multiple systems.  However, this approach relies on target domain sample trials to estimate the regression parameters, and is vulnerable if the trial domain is mismatched with the target domain. 

For efficient multi-domain learning, we can also use multiple domain datasets to learn a single network. Parameter sharing during training can be full or partial~\cite{Bilen2017,Rebuffi2018}. Since the task of each domain is the same,i.e. Arabic dialect identification, we will share all parameters for multi-domain learning with a single network. One of the advantages of multi-domain learning is that the input domain information is not needed whereas the single domain trained network needs domain information for maximum performance. 

Table~\ref{tab:baseline} shows DID accuracy on the MGB-3 and VarDial 2018 Test sets when using different domain training datasets. While systems trained using a single domain dataset such as $\mathcal{A}$ and $\mathcal{B}$ show robust performance only on the matched domain test set, system $\mathcal{A}$+$\mathcal{B}$ performs efficiently on both test set. Note that we doubled the number of filters in the neural network structure for system $\mathcal{A}$+$\mathcal{B}$ to match the network capacity. 

\begin{table}[ht]
\centering
\resizebox{0.48\textwidth}{!}{%
\begin{tabular}{cc|c|c}
\hline
\multirow{2}{*}{Training data} &\multirow{2}{*}{\begin{tabular}[c]{@{}c@{}}System\\ ID\end{tabular}}&\multicolumn{2}{c}{DID Accuracy (\%)} \\ \cline{3-4} 
 & & MGB-3 Test & VarDial 2018 Test \\ \hline
MGB-3 Train + MGB-3 Dev &$\mathcal{A}$ & \textbf{65.82} & 48.87 \\ \hline
YouTube Train &$\mathcal{B}$ & 51.27 & \textbf{86.40} \\ \hline
\begin{tabular}[c]{@{}c@{}}MGB-3 Train + MGB-3 Dev\\ + YouTube Train\end{tabular}& $\mathcal{A}+\mathcal{B}$ & 61.86 & 81.53 \\ \hline \hline
Fusion of $\mathcal{A}$ and $\mathcal{B}$ (optimized for $\mathcal{A}$)& - & \textbf{68.63} & 77.57 \\ \hline
Fusion of $\mathcal{A}$ and $\mathcal{B}$ (optimized for $\mathcal{B}$)& - & 57.84 & \textbf{86.94} \\ \hline
\end{tabular}%
}
\caption{Baseline dialect identification performance evaluation.}
\label{tab:baseline}
\end{table}

Score level fusion can be applied by logistic regression for maximum efficiency on multiple domains. The trials for optimizing parameters of logistic regression were generated by randomly combining utterances using the target domain training set.  The fusion approach achieves the best performance when the fusion rule was optimized for the target domain and achieves the worst performance on the test from another domain. Thus, the fusion approach is not practical if the system has no information about the domain. Although there is some performance degradation, the multi-domain trained system $\mathcal{A}$+$\mathcal{B}$ generally works on both domains.

\section{domain attentive fusion}
\subsection{Fusion layer for system combination}

\begin{figure}[htb]
    \centering
    \includegraphics[width=0.55\linewidth]{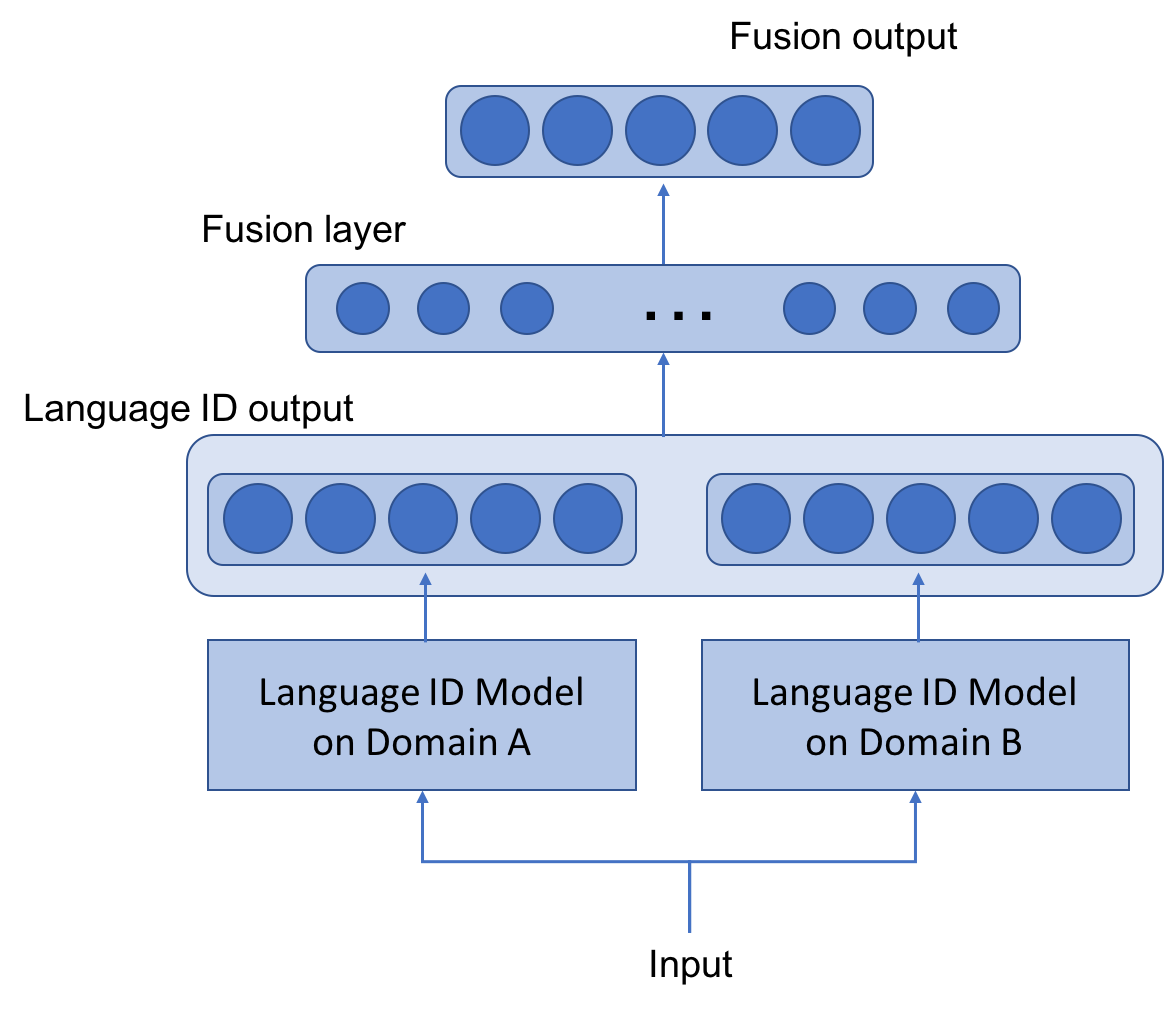}
    \caption{Fully-connected layer for score-level fusion.}
    \label{fig:fusionlayer}
\end{figure}

Traditional logistic regression for score-level fusion can be replaced by a neural network by adding a fusion layer on top of the single domain trained networks as shown in figure~\ref{fig:fusionlayer}. We used a fully connected layer with 600 hidden nodes and added a softmax layer which generates a score for each dialect. The fusion network was trained after the other networks were trained and fixed.  By learning from the training dataset, the fusion network dynamically selects the most useful scores which are invariant to domain mismatch.

\subsection{Self-attention based weighting}

\begin{figure}[htb]
    \centering
    \subfloat[Self-attention layer]{\includegraphics[width=0.62\linewidth]{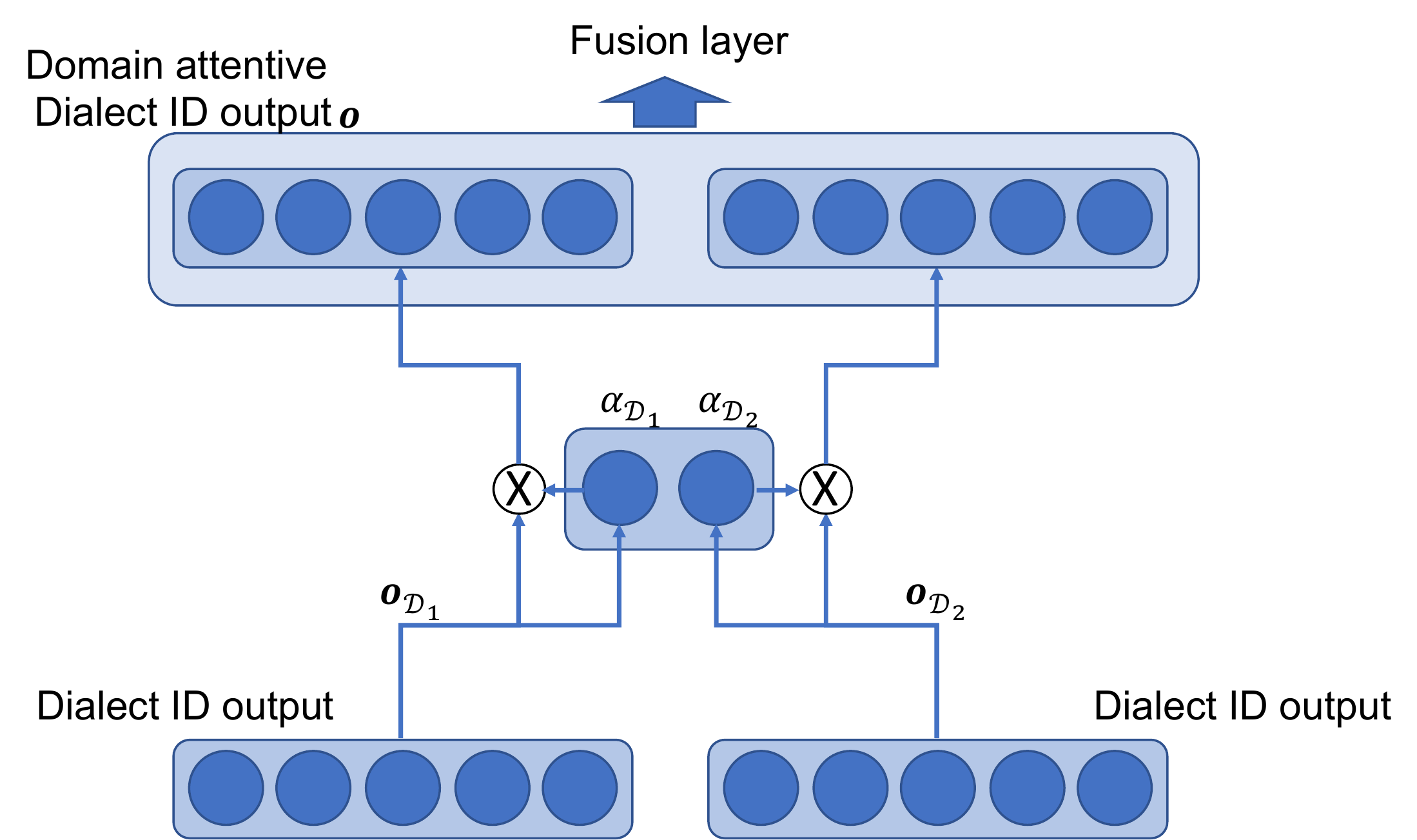}}
    
    \subfloat[Self-attention layer using embeddings]{\includegraphics[width=0.62\linewidth]{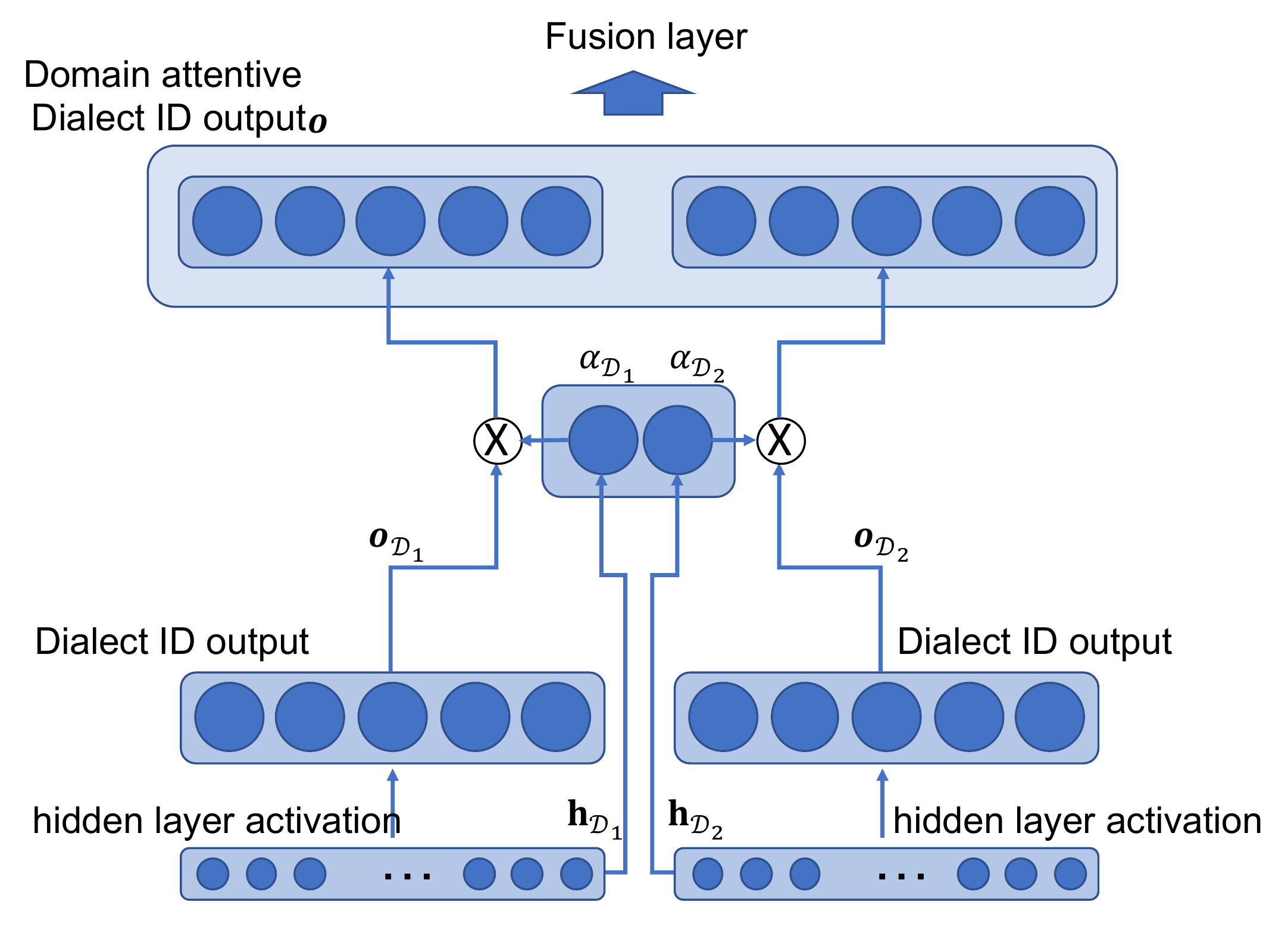}}
    \caption{Domain attentive fusion}
    \label{fig:attentive}
\end{figure}

The neural network attention mechanism is a powerful technique to focus on the significant or critical part of an input signal. An attention layer enables to focus on the important information of the input sequence by providing more weight on it.  Thus, in speech processing, it usually applied to the frame-level neural network layer to represent long sequence more effectively. In this research, we used an attention layer to adapt the domain by using multiple networks which were trained on different domains, as shown in Figure~\ref{fig:attentive}(a). 

Suppose the $L$-dimensional vector output $\textbf{o}_d$ of an end-to-end system trained using dataset with distribution $d\in\{\mathcal{D}_1, \mathcal{D}_2\}$ where $L$ is total number of dialects to be identified. We could learn a scalar score $e_d\in\mathbb{R}$ for output $\textbf{o}_d$ as 
\begin{equation}
\label{eq:attention_score}
e_d=f(\textbf{o}_d).
\end{equation}
The scoring function $f(\cdot)$ can be calculated as 
\begin{equation}
f(\textbf{o}_d) = \textbf{v}_d^T \tanh(\textbf{W}_d\textbf{o}_d+\textbf{b}_d)
\end{equation}
where $\textbf{W}_d$ is an $m$ by $L$ matrix and $\textbf{b}_d$ and $\textbf{v}_d$ are $m$-dimensional vectors. $m$ is a hyper-parameter that can be tuned. The normalized weights $\alpha_d$ can be computed as 
\begin{equation}
\alpha_d = \frac{exp(e_d)}{( exp(e_{\mathcal{D}_1}) + exp(e_{\mathcal{D}_2}) )}
\end{equation}
so, $\alpha_{\mathcal{D}_1} + \alpha_{\mathcal{D}_2}$ is equal to 1. Finally, we obtain the domain attentive output as 
\begin{equation}
\textbf{o} = [\alpha_{\mathcal{D}_1} * \textbf{o}_{\mathcal{D}_1} , \alpha_{\mathcal{D}_2} * \textbf{o}_{\mathcal{D}_2}]    
\end{equation}

Since domain related information is more likely remain in the intermediate layer than in the output layer, we can also incorporate hidden layer activations into the attention layer because they can be contain complimentary information when we calculate the end-to-end system output. In this case, we use the hidden layer activations $\textbf{h}_d$ as the input for the scoring function, and the scalar score can be calculated as
$e_d=f(\textbf{h}_d)$.  This approach is depicted in Figure~\ref{fig:attentive} (b).












\begin{table*}[t]
\centering
\resizebox{0.9\textwidth}{!}{%
\begin{tabular}{c|c|c|c|c|c|c|c|c|c}
\hlineB{2}
\multirow{3}{*}{Training data} & \multicolumn{9}{c}{Test on} \\ \cline{2-10} 
 & \multicolumn{3}{c|}{MGB-3 Test} & \multicolumn{3}{c|}{VarDial 2018 Test} & \multicolumn{3}{c}{Averaged} \\ \cline{2-10} 
 & Acc. & EER & Cavg & Acc. & EER & Cavg & Acc. & EER & Cavg \\ \hlineB{2}

MGB-3 Train + MGB-3 Dev ($\mathcal{A}$) & 65.82 & 20.43 & 19.60 & 48.87 & 28.39 & 28.50 & 58.35 & 24.41 & 24.05 \\ \hline
YouTube Train ($\mathcal{B}$) & 51.27 & 28.37 & 27.41 & 86.40 & 9.57 & 9.96  & 68.84 & 18.97 & 18.69 \\ \hline
MGB-3 Train + MGB-3 Dev + YouTube Train ($\mathcal{A}$+$\mathcal{B}$) & 61.86 & 22.92 & 21.41 & 81.53 & 11.13 & 11.76  &71.70 & 17.03 & \textbf{16.59} \\ \hline

Logistic regression fusion of $\mathcal{A}$ and $\mathcal{B}$ (optimized for $\mathcal{A}$)& \textbf{68.63} & \textbf{19.05} & \textbf{18.04} & 77.57 & 13.78 & 14.16  & \textbf{73.10}&	\textbf{16.42}	&16.10 \\ \hline
Logistic regression fusion of $\mathcal{A}$ and $\mathcal{B}$ (optimized for $\mathcal{B}$)& 57.84 & 24.36 & 23.35 & \textbf{86.94} & \textbf{9.23} & \textbf{9.56}  & 72.39&	16.80&	16.46 \\ \hline \hline

Using fusion layer on $\mathcal{A}$ and $\mathcal{B}$ (Figure~\ref{fig:fusionlayer})& 67.69 & 19.30 & 18.39 & 82.86 & 11.19 & 11.58 & 75.28 & 15.25 & 14.99 \\ \hline
Domain Attentive fusion of $\mathcal{A}$ and $\mathcal{B}$ (Figure~\ref{fig:attentive}~(a))& 67.49 & 18.52 & 18.01 & 83.93 & 10.03 & 10.22 & 75.71 & 14.28 & 14.12 \\ \hline
Domain Attentive fusion of $\mathcal{A}$ and $\mathcal{B}$ (Figure~\ref{fig:attentive}~(b))& \textbf{68.23} & \textbf{18.30} & \textbf{17.69} & \textbf{85.01} & \textbf{9.13} & \textbf{9.40} & \textbf{76.62} & \textbf{13.72} & \textbf{13.55} \\

\hlineB{2}
\end{tabular}%
}
\caption{Dialect identification performance for the ``Seen" test domain condition.}
\label{tab:seen}
\end{table*}

\section{Dialect Identification Experiments}
\label{sec:experiments}
For the end-to-end DID systems, we used MFCC features. To extract the features, a spectrogram was computed using a 400 sample FFT window length with 160 sample advance which is equivalent to 25ms window and 10ms frame-rate for 16kHz audio. A total of 40 coefficients were extracted and then normalized to have zero mean and unit variance. The end-to-end structure is the same as in~\cite{Shon2018odyssey}, with four CNN and two FF layers as described in Section 3. The stochastic gradient descent (SGD) learning rate was 0.001 with a decay every 50,000 mini-batches with a factor of 0.98. Rectified Linear Units (ReLUs) were used for activation nonlinearities.   For the attention layer, we set $m$ as 10.

Performance was measured in accuracy, Equal Error Rate (EER) and minimum decision cost function C\textsubscript{avg}*100.  Accuracy was measured by choosing the dialect with the maximum score for each test utterance. Minimum C\textsubscript{avg}
*100 was computed from hard decision errors and a fixed set of costs and priors from~\cite{lre15}.

Depending on the experimental condition, we used different datasets for training the network. Since we have three domains for training and two domains for testing from the MGB-3 and VarDial 2018 datasets, we could partition our experimental conditions into two categories, ``seen" and ``unseen" test domains.  For the seen test condition, we used a training dataset which is matched to the test domain, so that all test domains are already seen when the network is learning. For the unseen test condition, we excluded the training dataset which matched the test domain, so that the network could not learn about the test domain dataset distribution.

\begin{table*}[t]
\centering
\resizebox{0.9\textwidth}{!}{%
\begin{tabular}{c|c|c|c|c|c|c|c|c|c}
\hlineB{2}
\multirow{3}{*}{Training data} & \multicolumn{9}{c}{Test on} \\ \cline{2-10} 
 & \multicolumn{3}{c|}{MGB-3 Test (Unseen)} & \multicolumn{3}{c|}{VarDial 2018 Test (Seen)} & \multicolumn{3}{c}{Averaged} \\ \cline{2-10} 
 & Acc. & EER & Cavg & Acc. & EER & Cavg & Acc. & EER & Cavg \\ \hlineB{2}
MGB-3 Train ($\mathcal{C}$)  & 48.79 & 31.80 & 30.74 & 41.14 & 34.70 & 34.27 & 44.97 & 33.25 & 32.51 \\ \hline
YouTube Train ($\mathcal{B}$) & 51.27 & 28.37 & 27.41 & 86.40 & 9.57 & 9.96 &68.84 & 18.97 & 18.69 \\ \hline
MGB-3 Train + YouTube Train ($\mathcal{B}$+$\mathcal{C}$) & \textbf{56.37} & \textbf{25.07} & \textbf{24.10} & 83.85 & 9.87 & 10.30 & 70.11 & \textbf{17.47} &\textbf{ 17.20} \\ \hline
Logistic regression fusion of $\mathcal{B}$ and $\mathcal{C}$ (optimized for $\mathcal{C}$)& 55.29 & 25.67 & 24.84 & 83.26 & 11.09 & 11.15 & 69.28 & 18.38 & 18.00 \\ \hline
Logistic regression fusion of $\mathcal{B}$ and $\mathcal{C}$ (optimized for $\mathcal{B}$)& 54.22 & 26.69 & 25.67 & \textbf{87.56} & \textbf{8.96} & \textbf{9.36} & \textbf{70.89} & 17.83 & 17.52 \\ \hline \hline
Using fusion layer on $\mathcal{B}$ and $\mathcal{C}$ (Figure~\ref{fig:fusionlayer}) & 54.76 & 26.29 & 25.48 & 85.11 & 9.97 & 10.28 & 69.94 & 18.13 & 17.88 \\ \hline
Domain Attentive fusion of $\mathcal{B}$ and $\mathcal{C}$ (Figure~\ref{fig:attentive}~(a))& \textbf{55.83} & 25.67 & 24.92 & 85.63 & 9.84 & 9.97 & 70.73 & 17.76 & 17.45 \\ \hline
Domain Attentive fusion of $\mathcal{B}$ and $\mathcal{C}$ (Figure~\ref{fig:attentive}~(b))& 55.76 &\textbf{ 25.03} & \textbf{24.05} &\textbf{ 86.90} & \textbf{8.36} & \textbf{8.71} & \textbf{71.33} & \textbf{16.70 }& \textbf{16.38} \\

\hlineB{2}
\end{tabular}%
}
\caption{Dialect identification performance for the ``Unseen" and ``Seen'' test domain conditions.}
\label{tab:unseen}
\end{table*}

\subsection{Seen domain test condition results}

Table~\ref{tab:seen} shows the ``seen" condition experimental results whereby both the MGB-3 and VarDial 2018 domains could be learned in the training process.
From the results, we observe that training individual networks for each domain and fusing the results yields better performance than training multiple domains using a single network. 
It is interesting that the proposed domain attentive fusion perform remarkably better on both domains, and even better than logistic regression fusion which is optimized for each domain. Note that the domain attentive fusion approach doesn't need target domain information {\it a priori} and the attention layer automatically generates the weight of the network for fusion for an arbitrary input. 

\subsection{Unseen domain test condition results}
Table~\ref{tab:unseen} shows the ``unseen" test condition experimental results whereby only the VarDial2018 domain was learned by the end-to-end network, so that the MGB-3 domain was unseen in the training process.
On average, the domain attentive fusion approach shows the best performance among all approaches. Performance improvements on the unseen domain is not as impressive as for the ``seen" domain condition.  However, we verified that the attention layer still learns about domain information from the end-to-end system output and the hidden layer activations and generates a reasonable result. 

\section{Discussion}

For both ``seen" and ``unseen" conditions, the domain attentive fusion shows performance improvements compared to conventional approaches.  Apart from the performance, the proposed approach has a significant advantage when the input domain is unknown for practical situations. We believe that this approach can be extended to multiple domains and will enable the automatic calculation of the contribution of each sub-system to achieve the best result.

\begin{figure}[ht]
  \centering
  \subfloat[System $\mathcal{B}$\newline(Accuracy=86.40\%)]{\includegraphics[width=0.24\textwidth]{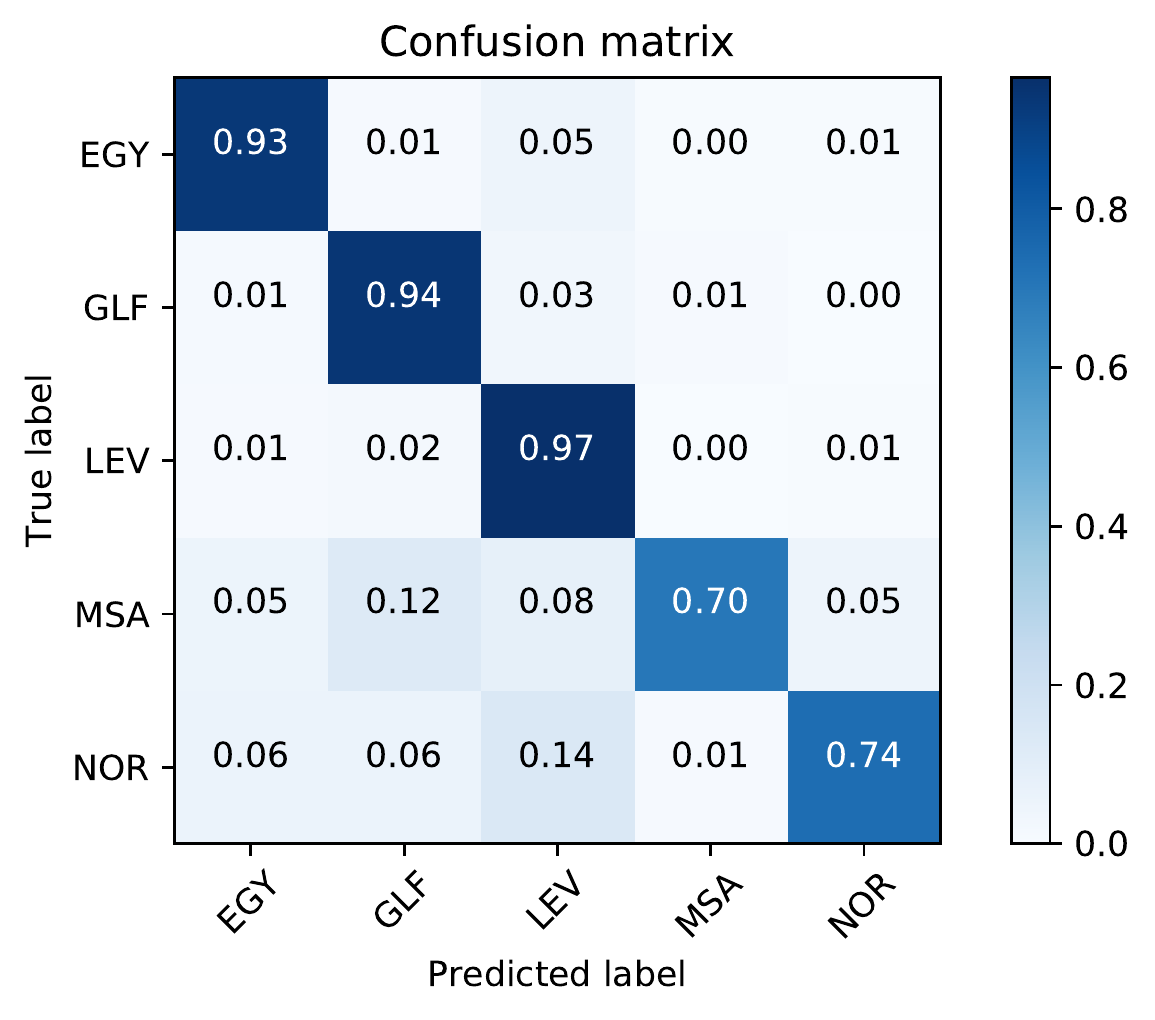}}
  \subfloat[After domain attentive fusion\newline
  (Accuracy=85.01\%)]
  {\includegraphics[width=0.24\textwidth]{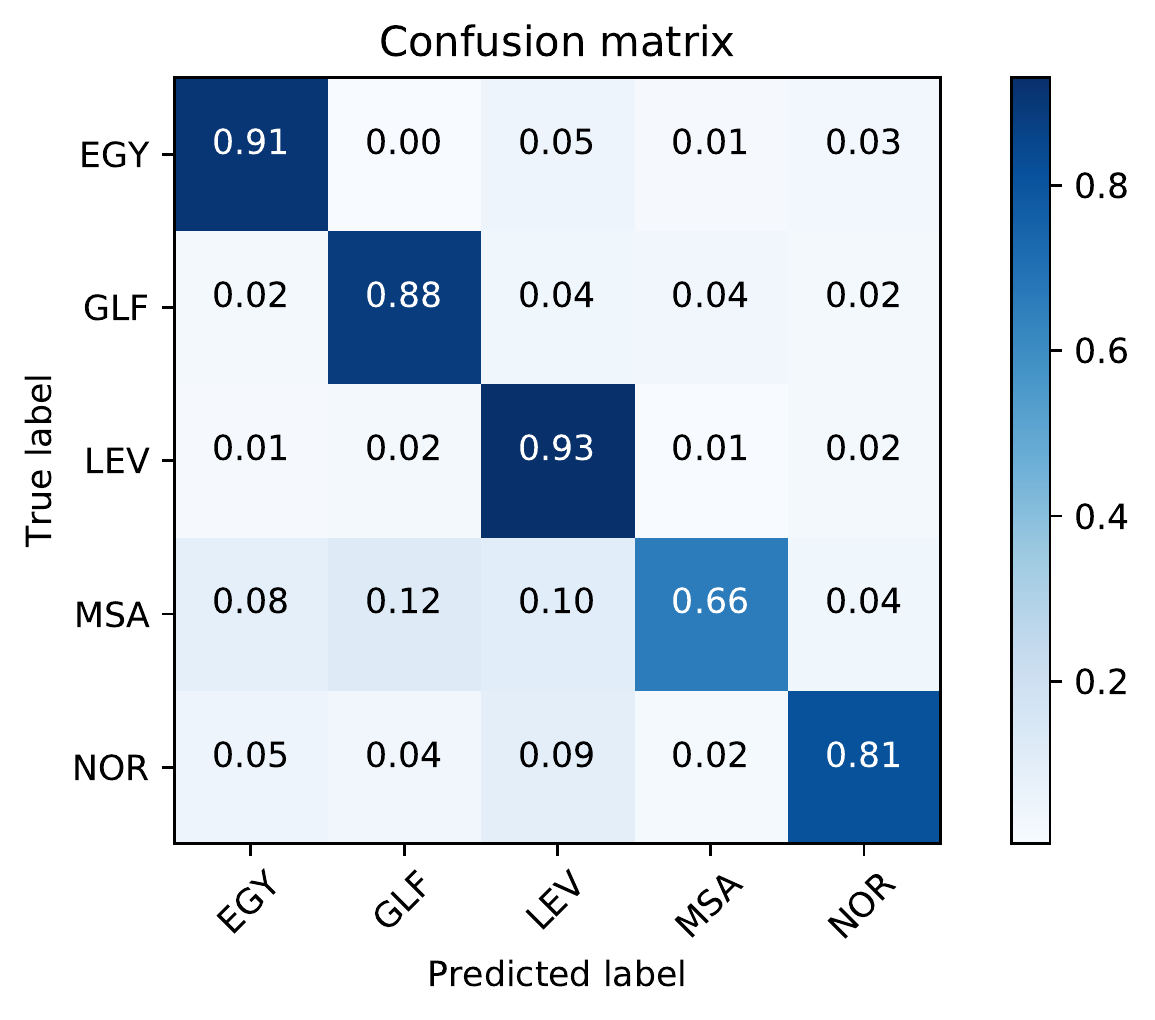}}
  \caption{DID confusion matrix on VarDial 2018 Test.}
\label{fig:confusion}
\end{figure}

Figure~\ref{fig:confusion} shows a confusion matrix of system $\mathcal{B}$ and the domain attentive fusion system in the last row of Table~\ref{tab:seen} on the VarDial 2018 test. Since the amount of data for each dialect is not balanced (see table~\ref{tab:data}, the confusion matrix shows poor performance of MSA and NOR compared to others.  We believe that this unbalanced situation always happens in low-resourced languages such as dialects, and we plan to address this issue in the future. 

\section{Conclusion}

A neural network-based end-to-end system has achieved the best performance on dialect/language identification tasks. But it remains vulnerable to domain mismatches, especially when the test domain is unknown. To recognize and adapt input domains automatically, we propose a domain attentive layer for fusion of multiple networks that are trained on a single domain. A domain attentive layer calculates the contribution of each network automatically by using the end-to-end language identification system outputs or hidden layer activations. The proposed approach was shown to be robust on test conditions without any {\it a priori} target domain knowledge.

For future work, we plan to expand the Arabic dialect identification task from 5 dialects to a larger number of country-specific dialects. We also plan to explore the scalability of the proposed approach to multiple domains, and balance performance using unbalanced datasets.

\clearpage
\newpage

\bibliographystyle{IEEEbib}
\bibliography{Template}

\end{document}